
\input phyzzx
\REF\REVIEW{For reviews of coherent states, see, e.g.,
J. R. Klauder ``A Coherent-State Primer" in {\it Coherent States.
Applicatons in Physics and Mathmatical Physics}, eds. J. R. Klauder
and B. S. Skagerstam (World Scientific, Singerpore, 1985);
A. Perelomov, {\it Generalized Coherent States and Their Applications}
(Springer-Verlag, Berlin, 1986).}
\REF\KLAUDER{J. R. Klauder, Phys. Rev. {\bf D19} (1979) 2349.}
\REF\KURATSUJIon{H. Kuratsuji and T. Suzuki, J. Math Phys. {\bf 21}
(1980) 472.}
\REF\LITTLEJOHN{R. G. Littlejohn, Phys. Rep. {\bf 138} (1986) 193,
and references therein.}
\REF\FUKUIon{T. Fukui, Prog. Theor. Phys. {\bf 87} (1992) 927.}
\REF\GENDENSTEIN{L. E. Gendenstein, JETP Lett. {\bf 38} (1983) 356;
L. E. Gendenstein and I. V. Krive, Sov. Phys. Usp. {\bf 28} (1985) 645.}
\REF\WITTEN{E. Witten, Nucl. Phys. {\bf B185} (1981) 513.}
\REF\INFELD{L. Infeld and T. E. Hull, Rev. Mod. Phys. {\bf 23} (1951)
21.}
\REF\SPIRIDONOV{V. Spiridonov, Phys. Rev. Lett. {\bf 69} (1992) 398.}
\REF\SUZUKI{T. Suzuki, Nucl. Phys. {\bf 398} (1983) 398.}
\REF\FUKUItw{T. Fukui, RCNP preprint 044.}

\def\yitp{Yukawa Institute for Theoretical Physics \break
Kyoto University, Kyoto 606-01, Japan}
\def\rcnp{Research Center for Nuclear Physics\break
Osaka University, Ibaraki, Osaka 567, Japan}
\def\refmark#1{[#1]}

\def\rket#1{|#1)}
\def\Re{{\rm Re}}
\def\Im{{\rm Im}}
\date{May, 1993} 
\titlepage
\title{\bf Shape-invariant potentials and an associated coherent state}
\vskip1.5cm\centerline{Takahiro Fukui}\address{\yitp}
\vskip1.5cm\centerline{N. Aizawa}\address{\rcnp}

\vskip2cm\abstract{
An algebraic treatment of shape-invariant potentials in
supersymmetric quantum mechanics is discussed. By introducing an
operator which reparametrizes wave functions, the
shape-invariance condition
can be related to a oscillator-like algebra. It makes
possible to define a coherent state associated with the shape-invariant
potentials. For a large class of such potentials, it is shown that the
introduced coherent state has the property of resolution of unity.
       }
\endpage
\vsize23.5cm 
Coherent states are one of important
concepts for physics today \refmark\REVIEW .
The original coherent state based on the Heisenberg-Weyl group
is extended for a number of Lie groups with square integrable
representations, and they have many applications in quantum mechanics.
In particular, they are used as bases of coherent state path
integrals \refmark{\KLAUDER ,\KURATSUJIon} or
dynamical wavepackets for describing the quantum
systems in semiclassical approximations \refmark\LITTLEJOHN .
Especially, if the hamiltonians have some dynamical symmetries,
they can be described easily and intuitively
by using the related coherent states as a basis.
This is quite remarkable
when we apply a semiclassical approximation to such systems;
higher order quantum corrections can be included in the classical
approximation \refmark{\FUKUIon}.
There may exist several different dynamical symmetries
if there exist several solvable potentials.
Therefore, we often select different coherent states for different
systems.
However, it was shown \refmark{\GENDENSTEIN} that a large class of
these solvable potentials are characterised by a simple property, i.e.,
a discrete reparametrization
invariance, called shape-invariance, introduced
by means of the supersymmetric quantum mechanics \refmark{\WITTEN}.
It is a generalization of the old factorization method \refmark\INFELD ,
in which possible six types of exactly solved potentials
are classified.
It is still an open question, however, to confirm the degree of its
generalization and to classify the possible shape-invariant
potentials.

Recently, Spiridonov analyzed a new class of
shape-invariant potentials \refmark\SPIRIDONOV .
It is not under a change of parameters $a\rightarrow f(a)$ but under
a dilatation of the dynamical varible $x\rightarrow qx$ that he demands,
as a shape-invariance condition,
the superpotentials are invariant.
Therefore, such potentials have a property of self-similarity.
Surprisingly, their dynamical symmetry is the quantum
algebra $su_q(1,1)$.
The key idea is to introduce the dilatation operator
$T_qf(x)=f(qx)$, which makes possible to express the condition of
self-similarity as the commutation relation of the q-oscillator,
and it enables us to construct the $ su_q(1,1) $ generators.
Then questions arise:
Are there any algebra-like structure in the usual shape-invariant
potentials as well as the self-similar ones?
If so, is it possible to introduce anything like a coherent state
describing a large class of such potentials simultaneously?

In this letter, a Lie algebra-like treatment of the shape-invariant
potentials is developed.
We first introduce an operator $T$ denoting reparametrization
$Tg(x,a)=g(x,a_1)$, where $a_1=f(a)$,
and then represent the shape-invariance condition
as a commutation relation. Based on this, we define a
coherent state associated with the shape-invariant potentials. It is
shown that this coherent state has the property of resolution of unity
for a large class of solvable potentials.

Let us introduce the superpotential $W(x,a)$
dependent on a parameter $a$ and assume that it satisfies the following
shape-invariance condition \refmark{\GENDENSTEIN}
$$
W^2(x,a)+W'(x,a)=W^2(x,a_1)-W'(x,a_1)+2R(a_1) ,
                                                                 \eqn\ON
$$
where $a_1=a-1\equiv f(a)$ and $R$ is a function independent of  $x$.
Strictly speaking, shape-invariant potentials are not exhausted
in this choice of parameters,
namely, it is possible, for example, to introduce
more than two parameters. This choice corresponds to the one discussed
in Ref.{\refmark\INFELD}.
However, we consider that factorization method in Ref.{\refmark\INFELD}
is a prototype of the shape-invariance condition, and in this letter,
we confine our attention to this case and
illustrate our basic ideas.

The shape-invariance condition {\ON} can be rewritten in the
following way. First, let $T$ be defined
$$
T\ket{\phi (x,a)}=\ket{\phi (x,a_1)}.
                                                                 \eqn\TW
$$
Namely, it is an operator which denotes the reparametrization of $a$.
{}From this definition, operators
$O(x,a)$, which depends on $a$, are transformed as
$$
O(x,a_1)\ket{\phi (x,a_1)}\equiv TO(x,a)\ket{\phi (x,a)}
=TO(x,a)T^{-1}\ket{\phi (x,a_1)},
$$
where $TT^{-1}=1$, i.e., $T^{-1}\ket{\phi (x,a)}=\ket{\phi
(x,a_{-1})}$ with $a_{-1}=a+1$.
Thus we have
$$
O(x,a_1)=TO(x,a)T^{-1}.
                                                                 \eqn\TH
$$
Next we define the hermitian conjugate of $T$. From the definition
{\TW}, we have the corresponding bra vector
$\bra{\phi (x,a)}T^\dagger =\bra{\phi (x,a_1)}$.
It should be noted here that the $ T$ does not preserve the
inner product in general;
$\langle\phi (x,a_1)|\varphi (x,a_1)\rangle\ne
\langle\phi (x,a)|\varphi (x,a)\rangle$.
We do not further enter into a mathematically rigorous
definition of $ T$,
rather, we use it as a convenient tool for developing an
algebraic treatment.

By using $T$, the shape-invariance condition {\ON} is written as
$$
D(a)D^\dagger (a)=T\{D^\dagger (a)D(a)+R(a)\}T^{-1},
                                                                 \eqn\FO
$$
where
$$
D(a)\equiv\{W(x,a)+\partial_x\}/\sqrt{2}.
                                                                 \eqn\FI
$$
Here and in the following, we do not denote explicitly $x$-dependence
of functions and operators.
With these, we define new operators \refmark{\SPIRIDONOV}
$$
\eqalign{
&A_+(a)=D^\dagger (a)T,\cr
&A_-(a)=T^{-1}D(a),\cr
        }                                                        \eqn\SI
$$
Then the shape-invariance condition {\FO} is simply given by
$$
[A_-(a), A_+(a)]=R(a),
                                                                 \eqn\SE
$$
Though this commutation relation resembles that of Heisenberg-Weyl
algebra, it differs in the following points; i) $A_-$ and $A_+$ are
not hermitian conjugate each other. ii) It is not a closed relation,
namely, $A_\pm$ and $R$ are not commutative in general.

With these preparations, we examine the properties of the eigenstates
of the hamiltonian $H_0=D^\dagger (a_0)D(a_0)$.
We assume it has a zero energy state, i.e., there exist a normalizable
state $\ket{\psi_0(a_0)}$ which satisfy
$$
D(a_0)\ket{\psi_0(a_0)}=A_-(a_0)\ket{\psi_0(a_0)}=0 .
                                                                 \eqn\EI
$$
In the coordinate representation it is given by
$\bra{x}\psi_0(a_0)\rangle\propto\exp[-\int^xW(y,a_0)dy]$, which
should be square integrable \refmark{\WITTEN}.
According to Gendenstein \refmark{\GENDENSTEIN},
we define the sequence of hamiltonians
$$
\eqalign{
&H_0\equiv A_+(a_0)A_-(a_0), \cr
&H_1\equiv A_-(a_1)A_+(a_1)=A_+(a_1)A_-(a_1)+R(a_1),\cr
&H_2\equiv A_-(a_2)A_+(a_2)+R(a_1)=A_+(a_2)A_-(a_2)+R(a_1)+R(a_2),\cr
&~~~~~~~~~~~~~~~~~~~~~~~~~~~~~~~~~~~~~~~~~~\vdots\cr
&H_n\equiv A_-(a_n)A_+(a_n)+\sum_{k=1}^{n-1}R(a_k)=
A_+(a_n)A_-(a_n)+\sum_{k=1}^nR(a_k),\cr
        }                                                        \eqn\NI
$$
where $a_k\equiv f(f(\cdots f(a_0)\cdots ))=a_0-k$.
For these hamiltonians, we can see that
$H_n$ and $H_{n+1}$ are superpartners, that is, they are isospectral
except for the ground state of $ H_n $.

Considering this property, we see that the eigenvalues and eigenstates
of $H_0$ are given by
$$
\eqalign{
&E_n(a_0)=\sum_{k=1}^nR(a_k) ,\cr
&\ket{\psi_n(a_0)}\propto \{A_+(a_0)\}^n\ket{\psi_0(a_0)} ,\cr
        }                                                      \eqn\ONZE
$$
provided there exist normalized states $\ket{\psi (a_k)}$ satisfying
$A_-(a_k)\ket{\psi (a_k)}=0$ for $k=0,...,n$ .
The last of Eq.{\ONZE} is easily verified by the commutation relation
$$
[H_0, \{A_+(a_0)\}^n]=\left\{\sum_{k=1}^nR(a_k)\right\} \{A_+(a_0)\}^n ,
$$
which shows  $H_0\{A_+(a_0)\}^n\ket{\psi_0(a_0)}=
(\{A_+(a_0)\}^nH_0+\sum_{k=1}^nR(a_k)\{A_+(a_0)\}^n)
\break\ket{\psi_0(a_0)}=
E_n(a_0)\{A_+(a_0)\}^n\ket{\psi_0(a_0)}$.
It should be noted here that excited states are generated by one
operator $A_+(a_0)$ and this makes possible
to construct a coherent state.
In terms of $D(a_k)$, the last equation can be rewritten as
a more familar form
$$
\ket{\psi_n(a_0)}\propto D^\dagger (a_0)D^\dagger (a_1)
\cdots D^\dagger (a_{n-1})\ket{\psi_0(a_n)}.
                                                               \eqn\ONON
$$
The level scheme is summarized in Figure.

Next consider the normalization of the eigenstates.
Let $\ket{\psi_n(a_k)}$ be the normalized eigenstates, then
$\ket{\psi_n(a_{k+1})}=T\ket{\psi_n(a_k)}$ with
$\langle\psi_n(a_{k+1})|\psi_n(a_{k+1})\rangle =1$. Therefore
$$
A_+(a_0)\ket{\psi_n(a_0)}=D^\dagger (a_0)\ket{\psi_n(a_1)}
=\sqrt{N_{n+1}(a_0)}\ket{\psi_{n+1}(a_0)} ,
$$
which shows
$$
\eqalign{
N_{n+1}(a_0)&=\bra{\psi_n(a_1)}D(a_0)D^\dagger (a_0)\ket{\psi_n(a_1)}\cr
&=\bra{\psi_n(a_1)}H_1+R(a_1)\ket{\psi_n(a_1)}
=\sum_{k=1}^{n+1}R(a_k).\cr}
$$
Thus we obtain
$$
\eqalign{
&A_+(a_0)\ket{\psi_n(a_0)}
=\sqrt{\sum_{k=1}^{n+1}R(a_k)}\ket{\psi_{n+1}(a_0)} ,\cr
&A_-(a_0)\ket{\psi_n(a_0)}
=\sqrt{\sum_{k=0}^{n-1}R(a_k)}\ket{\psi_{n-1}(a_0)} .\cr
         }                                                     \eqn\ONTW
$$
As a result, the normalized eigenstates of $H_0$ are given by
$$
\ket{\psi_n(a_0)}={1\over\sqrt{[n]_0!}}\{A_+(a_0)\}^n\ket{\psi_0(a_0)}
                                                               \eqn\ONTH
$$
with simplified notations
$$
\eqalign{
&[n]_k\equiv R(a_{k+1})+R(a_{k+2})+\cdots R(a_{k+n}) ,\cr
&\widehat{[n]}_k\equiv [n]_kT ,\cr
&[n]_k!\equiv\widehat{[n]}_k\widehat{[n-1]}_k\cdots\widehat{[1]}_k
\cdot T^{-n} .\cr
        }                                                      \eqn\ONFO
$$
If $R$ is constants, the spectra are equidistant, and $[n]_k$ and
$[n]_k!$ reduce to the usual natural number and the
factorial, respectively. In general, however, they depend on
the parameter $a$, and therefore more complicated situations occur.
For example, $[n]_k$ does not commute with $T$ in general:
$[T, [n]_k]=([n]_{k+1}-[n]_k)T=(R(a_{k+1+n})-R(a_{k+1}))T$.

As we obtain the normalized eigenstates, we proceed to construct
the coherent states, i.e., the eigenstates of $A_-(a_0)$.
We assume in the following that the systems under consideration
have infinite bound states.
Let $\exp_k(x)$ be a generalized exponential function
defined by the use of
the generalized factorial {\ONFO}
$$
\exp_k(x)\equiv\sum_{n=0}^\infty{1\over [n]_k!}x^n .
                                                               \eqn\ONFI
$$
Then we can define the following states dependent on a complex
parameter $z$ as
$$
\eqalign{
\rket{z,a_0}&\equiv\exp_0\{zA_+(a_0)\}\ket{\psi_0(a_0)} \cr
&=\sum_{n=0}^\infty{1\over\sqrt{[n]_0!}}z^n\ket{\psi_n(a_0)} ,\cr
         }                                                     \eqn\ONSI
$$
where round ket means an unnormalized state.
It may be needless to say that we can define similar coherent states
$\rket{z,a_k}$ by replacing $\exp_0\rightarrow\exp_k$.
If we operate $A_-(a_0)$ to the state \ONSI , we have
$$
\eqalign{
A_-(a_0)\rket{z,a_0}&=\sum_{n=0}^\infty{1\over\sqrt{[n]_{-1}!}}
z^nA_-(a_0)\ket{\psi_n(a_0)}\cr
&=\sum_{n=1}^\infty
{1\over\sqrt{[n]_{-1}T\cdot [n-1]_{-1}!\cdot T^{-1}}}
z^n\sqrt{[n]_{-1}}\ket{\psi_{n-1}(a_0)} \cr
&=\sum_{n=1}^\infty
{1\over\sqrt{T[n-1]_{-1}!T^{-1}}}z^n\ket{\psi_{n-1}(a_0)}
=z\rket{z,a_0}.\cr
        }                                                      \eqn\ONSE
$$
For a reference' sake, we mention that without using the generalized
exponential function we can formally express the coherent state as
$$
\rket{z,a_0}=\left[1-z\left\{{1\over H_0}A_+(a_0)\right\}\right]^{-1}
\ket{\psi_0(a_0)}.
                                                               \eqn\ONEI
$$
If $R(a)=1$, this state corresponds exactly to the usual boson
coherent state. The overlap and norm of this state are given by,
from Eq.{\ONSI}
$$
\eqalign{
&(z, a_0|z', a_0)=\exp_0(\bar zz') ,\cr
&{\cal N}_z(a_0)\equiv (z,a_0|z,a_0)=\exp_0(|z|^2) .\cr
        }                                                      \eqn\ONNI
$$
We see that the state {\ONSI} is non-orthogonal as the usual coherent
state.
This property plays a crucial role especially in coherent states
path integrals when evaluated in the semiclassical approximation
\refmark{\SUZUKI ,\FUKUItw}.
Using the norm {\ONNI}, we obtain the normalized state
$$
\ket{z,a_0}\equiv{1\over\sqrt{{\cal N}_z(a_0)}}\rket{z,a_0}.
                                                               \eqn\TWZE
$$
It should be noted that this normalized state has the different
eigenvalue from the unnormalized one ;
$$
    \sqrt{{{\cal N}_z(a_0) \over {\cal N}_z(a_{-1})}}\; z,     \eqn\TWON
$$
since the parameter in the normalization factor is changed by the
action of $ A_-(a_0) $.
Up to now, we have examined coherent ket-states. We here refer to the
conjugate ones. From $A_-(a_1)\rket{z,a_1}=
D(a_0)\rket{z,a_0}=z\rket{z,a_1}$, we have the corresponding
relation between coherent bra-states
$(z,a_0|D^\dagger (a_0)=(z,a_1|\bar z$, i.e.,
$(z,a_0|A_+(a_0)=(z,a_1|\bar zT$.

Next consider an important property of the coherent state, i.e.,
the completeness relation.
We follow the classification of Ref.{\refmark\INFELD}
and confine our attention to systems with only bound states.
Constants $a,b,c,d$ in this reference correspond to
$\alpha ,\beta ,\gamma ,\delta$, respectively.

\noindent\undertext{(I) Types (C) and (D).}  These are
the simplest cases among the shape-invariant potentials.
Superpotentials $W$ and $R$ in Eq.{\ON} are given by
$$
\eqalign{
&W(x,a)=\cases{(a+\delta )/x+\beta x /2 & (C),\cr
        \beta x+\delta & (D),\cr}\cr
&R(a)=\beta ,\cr
        }                                                      \eqn\TWTW
$$
where $\beta$ and $\delta$ are some real constants.
In these cases we can set $R(a)=1$ without loss of generality. Then
$$
[n]_k=n,\quad [n]_k!=n! ,
                                                               \eqn\TWTH
$$
are independent of $k$, namely, they are the usual natural numbers and
factorial, respectively.
The coherent state becomes, therefore,
$$
\eqalign{
&\rket{z, a_0}=\sum_{n=0}^\infty{z^n\over\sqrt{n!}}\ket{\psi_n(a_0)},\cr
&{\cal N}_z(a_0)=\exp(|z|^2) .
        }                                                      \eqn\TWFO
$$
For the measure
$$
d\mu (z)={1\over\pi}d\Re zd\Im z ,
                                                               \eqn\TWFI
$$
we have the following completeness relation
$$
\eqalign{
\int&d\mu (z)\ket{z,a_0}\bra{z,a_0} \cr
&=\sum_{m,n=0}^\infty{1\over\sqrt{m!n!}}\ket{\psi_m(a_0)}
\bra{\psi_n(a_0)}{1\over\pi}\int d\Re zd\Im z
\exp (-|z|^2)\bar z^mz^n \cr
&=\sum_{n=0}^\infty\ket{\psi_n(a_k)}\bra{\psi_n(a_k)}. \cr
        }                                                      \eqn\TWSI
$$

\noindent\undertext{(II) Types (A) and (B).}
Superpotentials and $R$ are given by
$$
\eqalign{
&W(x,a)=\cases{\alpha (a+\gamma )\cot\alpha (x+p)+
\delta /\sin\alpha (x+p) & (A),\cr
i\alpha (a+\gamma )+\delta\exp (-i\alpha x) &(B),\cr}\cr
&R(a)=-\alpha^2(a+\gamma +{1\over 2}) ,\cr
        }                                                     \eqn\TWSE
$$
where $\alpha$ is in general a real or pure imaginary constant
for type (A), while it is pure imaginary for type (B),
and  $\gamma ,\delta$ and $p$ are real constants.
However, as we assume that the hamiltonian has only bound states,
we confine our attention to the type (A) with real $\alpha$.
Note that in $W$ and $R$, $\gamma$ always appears as a form
$a+\gamma$. Therefore,
we can put $\gamma=0$ without loss of generality.
Furthermore, $R$ can reduce to $R(a)=-(a+1/2)$ by replacing
$A_\pm\rightarrow A_\pm/\alpha$. Finally, we can fix the various
parameters by choosing $a_0$ properly. Among them,
for example, if we choose $a_0=-1/2$, we have a simple expression
for the coherent state {\ONSI} as follows; first we have
$$
\eqalign{
&[n]_k={1\over 2}n(2k+n+1) ,\cr
&[n]_k!={1\over 2^n}n!{(2k+2n)!\over (2k+n)!}. \cr
        }                                                      \eqn\TWEI
$$
It should be noted the factorial depends on $k$, namely, on $a$.
For $k=0$, we have $[n]_0!=(2n)!/{2^n}$ and
$$
\eqalign{
&\rket{z,a_0}=\sum_{n=0}^\infty
{1\over \sqrt{(2n)!}}z^n\ket{\psi_n(a_0)} ,\cr
&{\cal N}_z(a_0)=\sum_{n=0}^\infty{|z|^{2n}\over (2n)!}=\cosh
(|z|) ,\cr
        }                                                      \eqn\TWNI
$$
where we replace $\sqrt{2}z\rightarrow z$. For the mesure
$$
d\mu_0(z)={1\over 2\pi}
{\cal N}_z(a_0){\exp (-|z|)\over |z|}d\Re zd\Im z
                                                               \eqn\THZE
$$
we have
$$
\eqalign{
\int&d\mu_0 (z)\ket{z,a_0}\bra{z,a_0}\cr
&=\sum_{m,n=0}^\infty{1\over\sqrt{(2m)!(2n)!}}
\ket{\psi_m(a_0)}\bra{\psi_n(a_0)}{1\over2\pi}
\int d\Re zd\Im z{\exp (-|z|)\over |z|}\bar z^mz^n \cr
&=\sum_{n=0}^\infty
\ket{\psi_n(a_0)}\bra{\psi_n(a_0)} .\cr
         }                                                     \eqn\THON
$$
We can obtain the similar relation
for $\rket{z,a_k}$ by transforming the last equation by $T^k$.

\noindent\undertext{(III) Types (E) and (F).}
Superpotentials and $R$ are given by
$$
\eqalign{
&W(x,a)=\cases{\alpha a\cot\alpha (x+p)+q/a & (E),\cr
a/x+q/a & (F),\cr}\cr
&R(a)=-\alpha^2(a+{1\over 2})+{q^2\over 2}\left\{{1\over (a+1)^2}
-{1\over a^2}\right\},\quad
(\alpha =0\quad {\rm for}\quad{\rm F}), \cr
        }                                                      \eqn\THTW
$$
where $\alpha$ is in general a real or pure imaginary constant,
and $q$ is a real constant.
Note that the potentials in type (F) always have continuous states
as well as bound states and those
in type (E) are able to have only bound
states for real $\alpha$. Therefore we restrict ourselves
only to this type. In this case
we may also construct the coherent state based on Eq.{\ONSI}
in the same way as the above.
In an alternative way, to make use of the discussion in (II),
we can consider another coherent state by using,
instead of $A_\pm$, newly defined operators,
$$
\widetilde A_+(a)\equiv A_+(a)-{q\over\sqrt{2}a}T,\quad
\widetilde A_-(a)\equiv A_-(a)-T^{-1}{q\over\sqrt{2}a},
                                                               \eqn\THTH
$$
which satisfy $[\widetilde A_-, \widetilde A_+]=\widetilde R$
with $\widetilde R(a)=-\alpha^2(a+1/2)$.
Using these operators,
we can define a coherent state in the same properties as
the category (II).

In summary,
by introducing an operator $T$ denoting a reparametrization
of $a$, we have represented the shape-invariance condition as a form of
a commutation relation, and by using it we have constructed a
coherent state associated with the shape-invariant potentials.
We expect this state should play a similar role as the usual and
generalized coherent states have been playing in various field
in modern physics.

Finally, we note that in types (C) and (D),
the commutation relation {\SE}
is closed and is considered a non-unitary realization of the
Heisenberg-Weyl algebra.
In types (A) and (B), we can also rewrite Eq.{\SE}
in a closed form as follows:
$$
[K_+(a), K_-(a)]=-2K_0(a), \quad [K_0(a), K_\pm(a)]=\pm K_\pm (a),
                                                               \eqn\THFO
$$
where
$$
K_+(a)=\sqrt{{2\over\alpha}}A_+(a), \quad
K_-(a)=\sqrt{{2\over\alpha}}A_-(a), \quad
K_0(a)={R(a)\over\alpha} .
                                                               \eqn\THFI
$$
Then we can consider that shape-invariance condition is a non-unitary
realization of the $su(1,1)$ algebra for the cases (A) and (B).
The hamiltonian is given simply by $H_0\propto K_+K_-$.
We see that the shape-invariant potentials have a quite simple
algebraic structure.
However, one should note that $K_\pm$ do not commute with
a function $f(a)$: $[K_\pm , f(a)]=(f(a_1)-f(a))K_\pm$.
Therefore, some unusual features may occur in the representation theory.
For example, we can construct a Casimir operator which commute with
all generators of Eq.{\THFI}. However, its eigenvalues are not
commutative with these generators since they depend on $a$.

It may be interesting to consider a possibility of
new shape-invariant potentials based on some other algebras.

\FIG\level{The level scheme}
\refout\vskip1cm
\centerline{{\fourteenrm FIGURE CAPTION}}
The level scheme
\bye